# Actively Controllable Topological Phase Transition in Phononic Beam Systems

Weijian Zhou[1], Weiqiu Chen[2], Michel Destrade[3] and C.W. Lim[1*]

[1]Department of Architecture and Civil Engineering, City University of Hong Kong, Tat Chee Avenue, Kowloon, Hong Kong SAR, P.R. China

[2]Department of Engineering Mechanics, Zhejiang University, Yuquan Campus, Hangzhou 310027, P.R. China

[3]School of Mathematics, Statistics and Applied Mathematics, NUI Galway, Ireland.

**Abstract**

Topological insulators, which allow edge or interface waves but forbid bulk waves, have revolutionized our scientific cognition of acoustic/elastic systems. Due to their nontrivial topological characteristics, edge (interface) waves are topologically protected against defects and disorders. This superior and unique characteristic could lead to a wealth of new opportunities in applications of quantum and acoustic/elastic information processing. However, current acoustic/elastic topological insulators are still at an infancy stage where the theory and prediction only work in laboratories and there are still many problems left open before promoting their practical applications. One of the apparent disadvantages is their narrow working frequency range, which is the main concern in this paper. We design a one-dimensional phononic beam system made of a homogeneous epoxy central beam sandwiched by two homogeneous piezoelectric beams, and covered with extremely thin electrodes, periodically and separately placed. These electrodes are connected to external electric circuits with negative capacitors. We show that a topological phase transition can be induced and tuned by changing the values of the negative capacitors. It follows that the working frequency of the topologically protected interface mode can be widely changed, such that the working frequency range of the topological insulator can be considerably 'broadened'. This intelligent topological device may also find wide applications in intelligent technologies that need controllable information processing of high precision.

---

[*]Corresponding author. E-mail: bccwlim@cityu.edu.hk

**List of symbols**

| | |
|---|---|
| $E_{PA}, E_{PB}$ | The effective Young modulus of the piezoelectric beam. |
| $C_{NA}, C_{NB}$ | Negative capacitances. |
| $C_{PA}^S, C_{PB}^S$ | Capacitances of the piezoelectric beams at constant strain. |
| $\varepsilon_{33}^S$ | Dielectric constant for constant strain. |
| $b$ | Width of the sandwiched beam. |
| $h_P$ | Thickness of the piezoelectric beam. |
| $l_A, l_B$ | Length of the piezoelectric beams. |
| $D_T$ | Bending stiffness of the sandwiched beam. |
| $\rho_T$ | Equivalent density per unit length of the sandwiched beam. |
| $D_{TA}$ | $D_{TA} = 2E_{PA}I_P + E_C I_C$. |
| $D_{TB}$ | $D_{TB} = 2E_{PB}I_P + E_C I_C$. |
| $\rho_C, \rho_P$ | Mass densities of the central beam and the piezoelectric beam. |
| $I_C, I_P$ | Moment of inertia with respect to the neutral line. |
| $g$ | $g = 2m\pi/L$ is the reciprocal-lattice constant, $m$ is an integer. |
| $Z_L$ | $Z_L = 1/sC_L$ is the impedance of a capacitor. |
| $C_P^\sigma$ | Capacitance of the piezoelectric beam at constant stress. |
| i | $\sqrt{-1}$ |
| $\varepsilon_{33}^\sigma$ | Dielectric constant for constant stress. |
| $w$ | Flexural wave deflection |
| $\beta$ | Capacitance parameter, $D_{TA} = (1-\beta)D_{T0}$, $D_{TB} = (1+\beta)D_{T0}$, |
| $E_{P0}$ | $E_{P0} = \frac{1}{s_{11}^E}\{1 + d_{31}^2 b l_n / [s_{11}^E h_p (C_{Nn} + C_{Pn}^S)]\}$. |
| $D_{T0}$ | Referential flexural rigidity, $D_{T0} = 2E_{P0}I_P + E_C I_C$. |
| $s_{11}^E$ | Compliance coefficient of the piezoelectric beam. |
| $d_{31}$ | Piezoelectric constant that couples mechanical deformation and electrical field of the piezoelectric beam. |
| $\gamma$ | Referential capacitance parameter. |
| $C_{N0}$ | Referential negative capacitance, $C_{N0} = \gamma C_{PB}^\sigma$. |
| $\beta_1, \beta_2, \beta_3, \beta_4$ | Defined in Eq. (18) |
| $\theta_n^{Zak}$ | Zak phase of $n$-th band |
| $\tilde{D}_{T0}$ | Dimensionless referential flexural rigidity, $\tilde{D}_{T0} = D_{T0}/D_0$. |
| $D_0$ | Flexural rigidity of the electrically open sandwiched beam |
| $u$ | Periodic-in-cell mode function, $u = e^{-ikL} w$ |
| $k$ | Bloch wave number |
| $L$ | Length of a unit cell |
| $\alpha$ | Displacement ratio, $\alpha = 20\log_{10}(|u_{output}|/|u_{input}|)$ |
| $I_1, I_2$ | Topological states |
| $\Omega_I$ | Stand for the two unit cells neighboring to the interface |
| $\Omega_T$ | Stand for the whole structure |
| $\eta$ | Quality factor to describe energy concentration around the interface |

# 1. Introduction

In recent years, the concepts in topological technologies such as the quantum Hall effect (QHE) [1, 2], quantum spin Hall effect (QSHE) [3-5] and quantum valley Hall effect (QVHE) [6, 7], have opened a new chapter of bosonic and classical waves with topology as the central principle. These concepts were initially proposed and studied in condensed matter physics [2] and later they were extended to optics, acoustics and mechanics [8-10]. One of the hallmarks of these concepts is the existence of topologically protected edge (interface) modes (TPEMs/TPIMs) [11]. They are 'topologically protected' in the sense that these edge/interface waves are tied to the topologically nontrivial nature of the underlying bands, and immune to backscattering from defects or disorders, making them robust against structural perturbations. Structures or materials supporting such TPEMs/TPIMs are also called as topological insulators (TIs) [12], which within a certain frequency range do not support bulk wave propagation but only allow edge/interface waves. These TIs have shown significant impact for applications in spintronics and quantum computation [13-15], high signal-to-noise ratio signal processing [16], energy tailoring [17], nondestructive testing, and wave-matter interaction for future quantum acoustics [18], to name a few.

QHE was first discovered in 1980 [19]. It opened a new field in condensed matter physics because it was the first time that a quantum state could be realized without spontaneously broken symmetry. QHE occurs when the time-reversal (TR) symmetry is broken. In electronic and photonic systems, the TR symmetry can be easily broken by the application of a magnetic field. However, in phononic systems, the materials are always passive and thus the TR symmetry is conserved. To mimic QHE, active components are needed in phononic systems. In recently proposed phononic analogies of QHE-based TIs, the TR symmetry was broken by insetting gyroscope [9, 20] and rotational flux [8] to bring angular momentum into the system. More recently, a new class of quantum states, known as the QSHE states, were theoretically predicted and experimentally observed [4, 5]. These quantum states were discovered by addressing the question: is it possible to mimic QHE-like states without applying external magnetic fields? In QSHE, the spin-orbit coupling [1, 21], which is universal to all materials, plays the role of the external magnetic fields. Since there is no application of external magnetic fields, the TR symmetry in QSHE states is conserved. A promising way to induce the QSHE in phononic systems is to break the space symmetry [22, 23], which is also an efficient way to induce QVHE. QVHE is analogous to QSHE: the valley index in QVHE plays a similar role of spin in QSHE, in mimicking the QHE-like states without breaking the TR symmetry [7]. Acoustic/elastic analogies of QSHE and QVHE have been a hot research topic and a comprehensive review of theoretical and experimental works on this field can be found in [24-32].

One of the main disadvantages of the existing phononic TIs is their apparent, narrow working frequency range. Particularly, in one-dimensional phononic systems, TPIM only exists at single frequency points [33-36]. These previous studies prove the existence of phononic TIs in principle, but they are very difficult to use in practical applications. This is because variation of environmental factors, i.e. environmental temperature [37], mechanical loading [38-40], external electric [41-43] and magnetic [44] fields, etc., may change the properties of these phononic TIs, and thus make their true working frequency shift out of the designed working range. Consequently, to make these TIs robust against perturbations of environmental conditions, we need to broaden their working frequency ranges, or to make these TIs respond automatically against environmental change to remain in their true working frequency range. Recently, Zhou et al. [45] proposed a membrane-type TIs that can be widely tuned by applying external voltage. Another motivation of our work is to develop intelligent phononic TIs with actively tunable working ranges or switchable functions, in order to meet different application requirements. Further, it was reported by Zhou et al. [46] that periodic electrical boundary conditions can be employed to induce and control the TPIM in a piezoelectric rod system. Motivated by these two reasons, in this study we aim to realize active control of TPIM over a wide range via external electric circuits.

An active one-dimensional phononic crystal (PC) is proposed in this paper. It is made of a homogeneous central epoxy central beam sandwiched by two piezoelectric beams. The surfaces of the piezoelectric beams are covered with separate and periodic silver electrodes. Between neighboring electrodes there exists a narrow insulating region (electrically open). These electrodes are connected to periodic electric circuits with step-wise negative capacitors (NCs), making each unit cell an A-B-A beam system. We show that the capacitance variation of NCs can induce topological phase transition and that TPIM can be realized in this way. The working frequency of TPIM can be smoothly and actively changed in a wide range by varying the NCs. During the controlling process the concentration performance of TPIM can be changed or maintained, depending on how these NCs are changed.

## 2. Modeling and basic equations

The geometrical phase inversion of acoustic waves in a phononic tube has been studied by Xiao et al. [33] and later its elastic counterparts in elastic beams were reported by Yin el al. [34]. However, TPIM, which is the most interesting phenomenon in these topological phononic systems, was found to work only at a single frequency point. Here, we propose a model to achieve geometric phase inversion by employing periodic electrical boundary conditions. In this way, the working frequency of TPIM can be tuned smoothly in a wide frequency range such that the working frequency domain of the TPIM can be significantly 'broadened'.

The composite 1D beam system considered here is displayed in Fig. 1. It is structured by an epoxy central beam sandwiched by two thin piezoelectric beams made of PZT-5H. Between each piezoelectric beam and the central beam, there is an extremely thin silver electrode layer and its electrical boundary condition (EBC) is set as ground. The top and bottom surfaces are periodically covered with extremely thin silver electrodes. These surface electrodes are represented as brown and green rectangles in Fig. 1. Between each two neighboring electrodes there is a narrow gray gap that is electrically open, playing the role of an insulator. The brown electrodes on the top and bottom surfaces are linked to NCs of capacitance $C_{NA}$ while the green electrodes are connected to NCs of capacitance $C_{NB}$. Thus, each unit cell can be seen as an A-B-A beam.

Marconi et al. [47] recently presented a similar structure, by using an experimental approach. They adopted the time-varying NCs to break the TR symmetry in an effort to realize non-reciprocal wave propagation. In contrast to that previous study, we mainly concentrate on the tunable TPIMs and reversed TR symmetry.

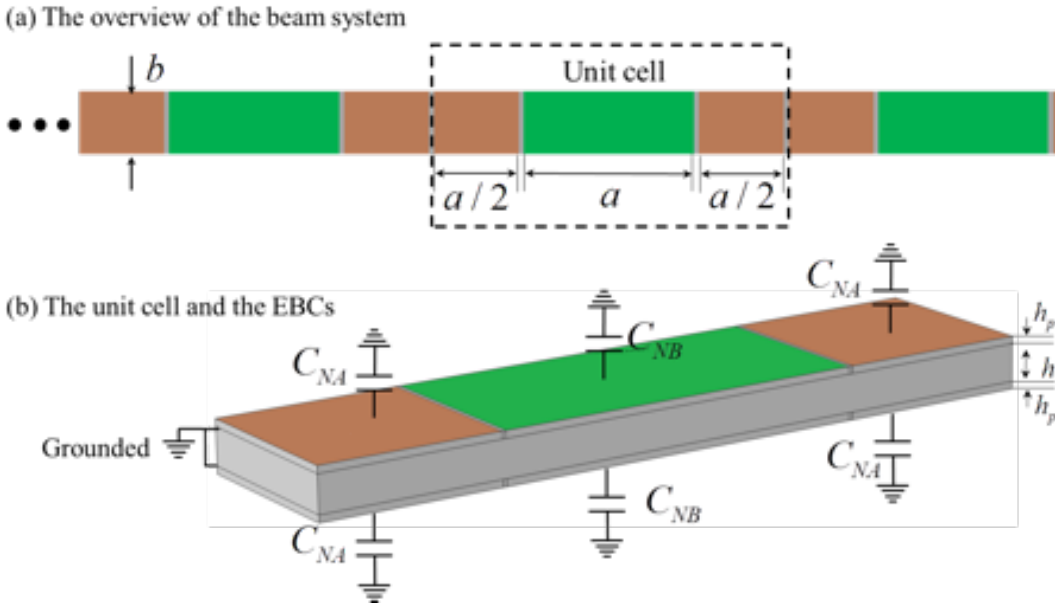

**Fig. 1**. A sandwiched beam system. (a) Overview of the beam; and (b) the unit cell and EBCs, with the central beam thickness $h$, piezoelectric beam thickness $h_p$, beam width $b$, brown rectangle with length $a/2$ and green rectangle with length $a$.

## 2.1 Dispersion relation

The piezoelectric beams are assumed perfectly bonded to the central beam. The effect of NC can be described by its influence on $E_P$, the effective Young modulus of the piezoelectric beam, which is a function of the capacitance [48, 49],

$$E_{P\alpha} = \frac{1}{s_{11}^E}\left[1 + \frac{d_{31}^2 b l_\alpha}{s_{11}^E h_P\left(C_{N\alpha} + C_{P\alpha}^S\right)}\right], \tag{1}$$

where the subscript $\alpha = A$ or $B$ stands for the A or B sub-beam, $s_{11}^E$ is the compliance coefficient, $d_{31}$ is a piezoelectric constant that couples mechanical deformation and electrical field of the piezoelectric beam, $C_{N\alpha}$ is the capacitance of NC, $C_{PA}^S = ab\varepsilon_{33}^S/2h_P$ and $C_{PB}^S = 2C_{PA}^S$ are the capacitances of the piezoelectric beams at constant strain, $\varepsilon_{33}^S$ is the dielectric constant for constant strain, $b$ is the width of the sandwiched beam, $h_P$ is the thickness of the piezoelectric beam, and $l_\alpha$ is the length of the piezoelectric beam, with $l_A = a/2$ and $l_B = a$. Thus, in the following analysis for flexural wave propagation, these piezoelectric beams will be treated as elastic beams, except for the Young's moduli that are functions of NCs, as expressed in Eq. (1).

The Euler-Bernoulli beam theory is adopted to model the sandwiched beam with rotational inertia of the cross section and shearing deformation neglected. The flexural wave deflection $w(x,t)$ is governed by the equation

$$\frac{\partial^2}{\partial x^2}\left[D_T(x)\frac{\partial^2}{\partial x^2}w(x,t)\right] + \rho_T\frac{\partial^2}{\partial t^2}w(x,t) = 0 \tag{2}$$

where $D_T(x)$ is the bending stiffness of the sandwiched beam and $\rho_T(x)$ is its equivalent density per unit length. To fulfill strain compatibility and Euler-Bernoulli beam assumptions, the bending stiffness $D_T(x)$ and the equivalent density $\rho_T$ can be expressed as

$$D_T(x) = \begin{cases} D_{TB}, & \text{if } |x - nL| \leq a/2, \\ D_{TA}, & \text{if } a/2 < |x - nL| \leq a, \end{cases} \tag{3}$$

$$\rho_T = 2\rho_P b h_P + \rho_C b h_C$$

where $D_{TA} = 2E_{PA}I_P + E_C I_C$, $D_{TB} = 2E_{PB}I_P + E_C I_C$, in which subscripts $C$ and $P$ denote parameters of the central beam and the piezoelectric beams, respectively, $\rho_C$ and $\rho_P$ are the mass densities, and $I_C$ and $I_P$ are the moment of inertia with respect to the neutral line, expressed as

$$I_C = \frac{1}{12}h_C^3 b, \quad I_P = \left(\frac{1}{4}h_C^2 h_P + \frac{1}{2}h_C h_P^2 + \frac{1}{3}h_P^3\right)b. \tag{4}$$

The plane wave expansion method [50] is adopted to consider flexural wave propagation in the beam system. Due to periodicity, $D_T(x)$ in Eq. (2) can be expressed in Fourier expansions as

$$D_T(x) = \sum_g \hat{D}_T(g)e^{igx} \tag{5}$$

where $g = 2m\pi/L$ is the reciprocal-lattice constant, $m$ is an integer, $i = \sqrt{-1}$, and the Fourier parameter $D_T(g)$ is expressed as

$$\hat{D}_T(g) = \begin{cases} \frac{1}{2}(D_{TA} + D_{TB}), & (g = 0), \\ \frac{1}{ga}[D_{TB} - D_{TA}]\sin\left(\frac{1}{2}ag\right), & (g \neq 0). \end{cases} \tag{6}$$

The solution of Eq. (2) can be expressed as

$$w(x,t) = e^{i\omega t} \sum_g \hat{w}(g) e^{i(k+g)x} \quad (7)$$

where $k$ is the Bloch parameter.

Substituting Eqs. (5) and (7) into Eq. (2), selecting $m = -M,...,0,...M$ to truncate the infinite summation (in total there are $N = 2M+1$ plane waves), and further using orthogonal relations of the exponential function, the following eigenvalue equation can be obtained as

$$\mathbf{K}\boldsymbol{\xi} - \rho_T \omega^2 \boldsymbol{\xi} = \mathbf{0} \quad (8)$$

where $K_{IJ} = (k+g_I)^2 \hat{D}_T(g_I - g'_J)(k+g'_J)^2$, $(I,J = 1,2,...N)$ and $\xi_J = \hat{w}(g_J)$. For a given Bloch parameter $k$, the frequency of the $n$-th branch can be determined by selecting the $n$-th eigenvalue from Eq. (8). Then the dispersion structure of the flexural waves in the beam system can be obtained by sweeping the Bloch parameter in the first Brillouin zone.

## 2.2 The stability consideration of NC

The NC is a mathematical model which can be realized by an artificial electric circuit [51], as shown in Fig. 2. The effective impedance of the negative impedance converter is

$$Z_{in} = -\frac{R_2}{R_1} Z_L \quad (9)$$

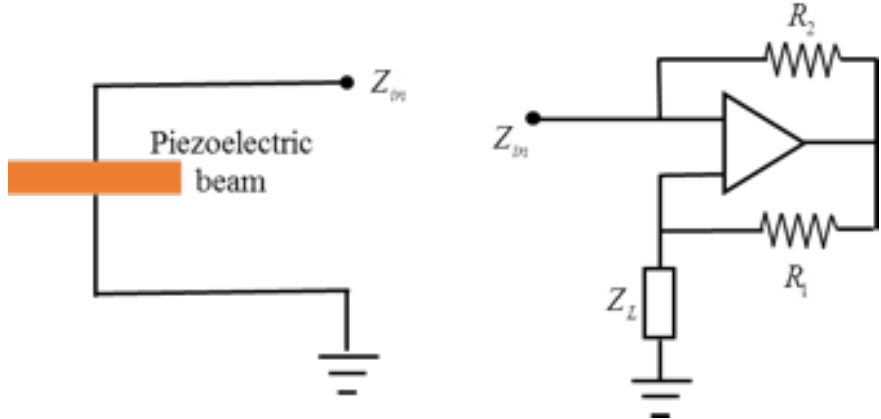

Fig. 2. The electric circuit to realize NC.

where $Z_L = 1/sC_L$ is the impedance of a capacitor. The effective impedance becomes

$$Z_{in} = \frac{1}{s \frac{-R_1 C_L}{R_2}} \quad (10)$$

From this expression, the electric circuit can be viewed as a NC with its negative capacitance expressed as

$$C_N = -\frac{R_1 C_L}{R_2} \quad (11)$$

Because a NC connected to the piezoelectric beam may induce electrical instability [51], we now consider this issue in the active system carefully. On the one hand, it is reported that a NC connected to piezoelectric beam in series (see Fig. 2) is stable if $|C_N| > C_P^\sigma$, with $C_P^\sigma$ being the capacitance of the piezoelectric beam at constant stress [51]. Here, the effective capacitances of sub-beams A and B at constant stress are expressed as $C_{PA}^\sigma = ab\varepsilon_{33}^\sigma/2h_P$ and $C_{PB}^\sigma = ab\varepsilon_{33}^\sigma/h_P$, with $\varepsilon_{33}^\sigma = \varepsilon_{33}^S + d_{31}^2/s_{11}^E$. On the other hand, a NC can achieve a very good control effect only when the value of NC is close enough to its instability boundary, i.e. $|C_N| \to C_P^\sigma$. However, in practice, any variation in environmental temperature will change the effective capacitance of the piezoelectric beam. Therefore, for a NC that is too close to its instability boundary, a change in the effective capacitance of the piezoelectric beam caused by temperature variation may result in electrical instability. Consequently, taking both stability and efficiency into consideration, a negative capacitance that is slightly away from its instability boundary is considered here. Generally, the absolute value of the negative capacitance should be at least 2% greater than $C_P^\sigma$ [51]; hence the system can be robust to small perturbations of environmental temperature and ensure high efficiency at the same time. Therefore, in this paper we set the negative capacitance in the range $|C_N| \geq 1.02 C_P^\sigma$.

For parametric analyses of topological phase transition in this study, we restrict attention to the two NC groups that obey the following relation

$$D_{TA} = (1-\beta)D_{T0}, \quad D_{TB} = (1+\beta)D_{T0}, \tag{12}$$

where $\beta$ is a defined capacitance parameter and $D_{T0} = 2E_{P0}I_P + E_C I_C$ is a referential flexural rigidity with $E_{P0}$ expressed as

$$E_{P0} = \frac{1}{s_{11}^E}\left[1 + \frac{d_{31}^2 b l_2}{s_{11}^E h_P (C_{N0} + C_{PB}^S)}\right], \tag{13}$$

in which $C_{N0} = \gamma C_{PB}^\sigma$, $(\gamma < -1.02)$ is a given referential negative capacitance, with $\gamma$ being a referential capacitance parameter. The referential flexural rigidity must be positive, i.e. $D_{T0} > 0$, or the system will be unstable. From this restriction we have the relation $h_P s_{11}^E (2I_P + E_C I_C s_{11}^E)(C_{N0} + C_{PB}^S) + 2abd_{31}^2 I_P < 0$, or

$$C_{N0} < -\frac{2abd_{31}^2 I_P}{h_P s_{11}^E (2I_P + E_C I_C s_{11}^E)} - C_{PB}^S. \tag{14}$$

Substituting Eqs. (1), (3) and (13) into Eq. (12) yields the expressions of $C_{NA}$ and $C_{NB}$ as

$$C_{NA} = \frac{abI_P d_{31}^2 (C_{N0} + C_{PB}^S)}{-\beta h_P s_{11}^E (2I_P + E_C I_C s_{11}^E)(C_{N0} + C_{PB}^S) + 2abd_{31}^2 I_P (1-\beta)} - C_{PA}^S,$$

$$C_{NB} = \frac{2abI_P d_{31}^2 (C_{N0} + C_{PB}^S)}{\beta h_P s_{11}^E (2I_P + E_C I_C s_{11}^E)(C_{N0} + C_{PB}^S) + 2abd_{31}^2 I_P (1+\beta)} - C_{PB}^S. \tag{15}$$

Since $C_{NA} < -1.02 C_{PA}^\sigma$ and $C_{NB} < -1.02 C_{PB}^\sigma$, we can have the following constraint via Eq. (15) as

$$\frac{abI_P d_{31}^2 (C_{N0} + C_{PB}^S)}{-\beta\left[h_P s_{11}^E (2I_P + E_C I_C s_{11}^E)(C_{N0} + C_{PB}^S) + 2abd_{31}^2 I_P\right] + 2abd_{31}^2 I_P} < C_{PA}^S - 1.02 C_{PA}^\sigma$$

$$\frac{2abI_P d_{31}^2 (C_{N0} + C_{PB}^S)}{\beta\left[h_P s_{11}^E (2I_P + E_C I_C s_{11}^E)(C_{N0} + C_{PB}^S) + 2abd_{31}^2 I_P\right] + 2abd_{31}^2 I_P} < C_{PB}^S - 1.02 C_{PB}^\sigma \tag{16}$$

Since $C_{PA}^\sigma \geq C_{PA}^S$ and $C_{PB}^\sigma \geq C_{PB}^S$, we find that right sides of the above inequalities (i.e. $C_{PA}^S - 1.02 C_{PA}^\sigma$ and $C_{PB}^S - 1.02 C_{PB}^\sigma$) are both negative. Furthermore, $C_{N0} < -1.02 C_{PB}^\sigma$. Therefore, it is obvious that the dominators in inequalities (16) must be positive and thus

$$-\beta h_P s_{11}^E (2I_P + E_C I_C s_{11}^E)(C_{N0} + C_{PB}^S) + 2abd_{31}^2 I_P (1-\beta) > 0,$$

$$\beta h_P s_{11}^E (2I_P + E_C I_C s_{11}^E)(C_{N0} + C_{PB}^S) + 2abd_{31}^2 I_P (1+\beta) > 0. \tag{17}$$

The inequalities (16) and (17) are satisfied if $\beta_a < \beta < \beta_b$, where $\beta_a = \max\{\beta_1, \beta_3\}$ and $\beta_b = \min\{\beta_2, \beta_4\}$, with

$$\beta_1 = \frac{2abd_{31}^2 I_P}{\left[h_P s_{11}^E (2I_P + E_C I_C s_{11}^E)(C_{N0} + C_{PB}^S) + 2abd_{31}^2 I_P\right]},$$

$$\beta_2 = \frac{-2abd_{31}^2 I_P}{h_P s_{11}^E (2I_P + E_C I_C s_{11}^E)(C_{N0} + C_{PB}^S) + 2abd_{31}^2 I_P},$$

$$\beta_3 = \frac{2abI_P d_{31}^2 (C_{N0} + 1.02 C_{PB}^\sigma)}{\left[h_P s_{11}^E (2I_P + E_C I_C s_{11}^E)(C_{N0} + C_{PB}^S) + 2abd_{31}^2 I_P\right](C_{PB}^S - 1.02 C_{PB}^\sigma)},$$

$$\beta_4 = \frac{2abI_P d_{31}^2 (C_{N0} + 1.02 C_{PB}^\sigma)}{-\left[h_P s_{11}^E (2I_P + E_C I_C s_{11}^E)(C_{N0} + C_{PB}^S) + 2abd_{31}^2 I_P\right](C_{PB}^S - 1.02 C_{PB}^\sigma)}.$$

(18)

The discussion above is based on electric stability analysis while, in addition, mechanical stability should also be considered, i.e. $D_{T1} > 0$ and $D_{T2} > 0$, which means $-1 < \beta < 1$. Therefore, the parameter $\beta$ should be in the range $\beta_{\min} < \beta < \beta_{\max}$, with $\beta_{\min} = \max\{\beta_a, -1\}$ and $\beta_{\max} = \min\{\beta_b, 1\}$.

## 3. Numerical simulations

In this section, a numerical analysis on the band structures and topological performance of the PC beam system is presented and the active control of TPIM is explored. The structural parameters of the beam system are set as $a = 2$ cm, $b = 1$ cm, $h_C = 0.2$ cm and $h_p = 0.1$ cm. The physical parameters of epoxy and PZT-5H are given in Table 1.

Table 1. The material parameters [48].

| Material properties | $\rho$ $(\text{kg}/\text{m}^2)$ | $E_{Young}$ $(10^{10} \text{ N}/\text{m}^2)$ | $s_{11}^E$ $(10^{-12} \text{ m}^2/\text{N})$ | $d_{31}$ $(10^{-10} \text{ C}/\text{N})$ | $\varepsilon_{11}^\sigma$ $(10^{-8} \text{ F}/\text{m})$ |
|---|---|---|---|---|---|
| Epoxy | 1142 | 0.41 | – | – | – |
| PZT-5H | 7500 | – | 16.50 | −2.74 | 3.01 |

The stability of the active system is considered first. The map of the allowed range of $\beta$ as a function of $\gamma$ is shown in Fig. 3(a). It can be seen that the allowed range gets wider when $\gamma$ increases from $\gamma = 1.02$ to $\gamma = \gamma_{cr}$, where the width of the allowed range reaches a maximum value. Subsequently, the width of the allowed range decreases as $\gamma$ further increases. Note that a wide allowed range of $\beta$ can achieve a wide controlling range of flexural rigidity $D_{TA}$ and $D_{TB}$, as well as a large mismatch between sub-beams A and B. Thus, to realize good tunability as well as wide band gaps, $\gamma$ should be close to $\gamma_{cr}$. In Fig. 3 the relation between $\gamma$ and the dimensionless referential flexural rigidity $\tilde{D}_{T0}$ is presented, where $\tilde{D}_{T0} = D_{T0} / D_0$, with $D_0$ being the flexural rigidity of the electrically open sandwiched beam. From this figure we know that NC can significantly change the flexural rigidity of the sandwiched beam, especially when NC is close to the instability boundary, i.e. when $\gamma \to 1.02$. This phenomenon agrees well with the discussion in the previous section.

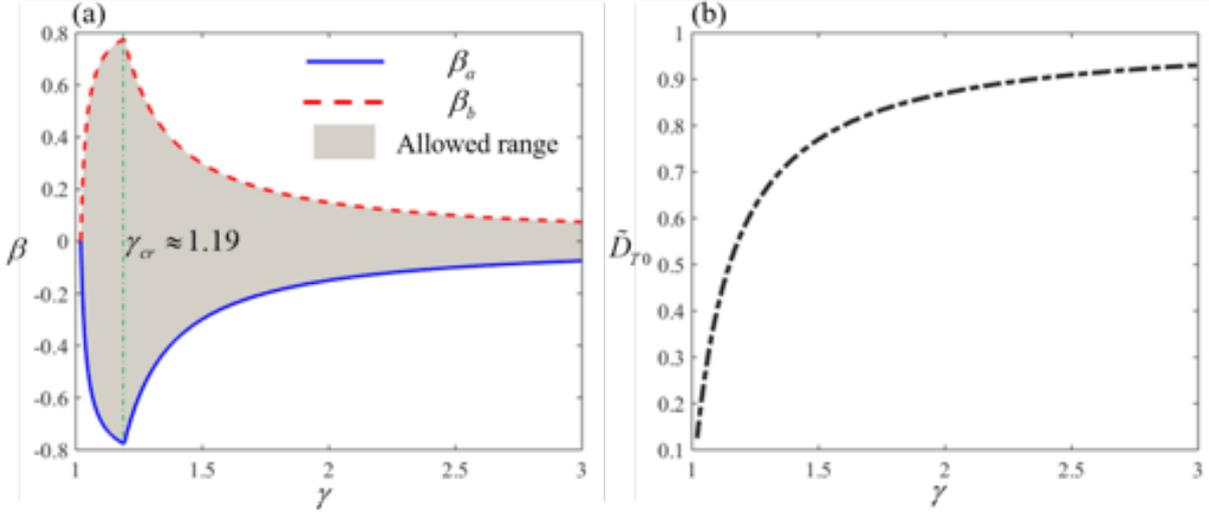

Fig. 3. The relation between $\gamma$, the allowed range of $\beta$ and $\tilde{D}_{T0}$.

In the following section, we set $\gamma = 1.19$ and investigate the tunable band structure of the active PC beam. For $\beta = 0$ ($C_{NA} = -1.19 C_{PA}^{\sigma}$ and $C_{NB} = -1.19 C_{PB}^{\sigma}$), we have $D_{TA} = D_{TB}$ and therefore the PC becomes homogeneous along the $x$-direction. For this reason, the dispersion curve shown in Fig. 4(a) is continuous and there is no band gap. In the figure, the dimensionless Bloch parameter $kL$ is swept in the whole first Brillouin zone $0 \leq kL \leq 2\pi$. A Dirac cone in the band structure located at $kL = \pi$ is observed. For $\beta = 0.6$ ($C_{NA} = -1.0419 C_{PA}^{\sigma}$ and $C_{NB} = -1.19 C_{PB}^{\sigma}$, the negative capacitance will not induce electrical instability), the 'homogeneous' PC beam system is no longer homogeneous (in this situation, $D_{TB} = 4 D_{TA}$). As a result, the homogeneous dispersion curve is split, and a band gap occurs.

To validate this new theoretical model (for all the following simulations, 41 plane waves are used), the numerical solutions are obtained by COMSOL Multiphysics 5.4, using 3D elements. A comparison in Fig. 4 between theoretical and numerical solutions shows that, on the one hand the physical phenomena of Dirac cone and band gap can be seen in both results; while on the other hand, the difference between numerical and theoretical dispersion curves is quite notable, especially in the high-frequency domain. Such a difference is expected. In the theoretical analysis, the Euler-Bernoulli beam model was employed to study the sandwiched beam system, where the shearing displacement and inertia are ignored. Furthermore, the displacement function along the thickness direction was constrained. These assumptions (or constraints) simplify the mathematical model but they also result in a stiffer beam system. This effect is very weak for a slender beam with low frequency. It is noted that the beam considered here is relatively thick because the length-to-thickness ratio of a unit cell is only 5. As a result, the theoretical beam model is stiffer than the numerical beam model and hence the theoretical dispersion curves are consequently higher than the numerical dispersion curves. Despite the relatively large difference between theory and numerical simulation, the theoretical analysis is still reliable and valuable because it captures the main physical phenomena and offers an explanation to the mechanism.

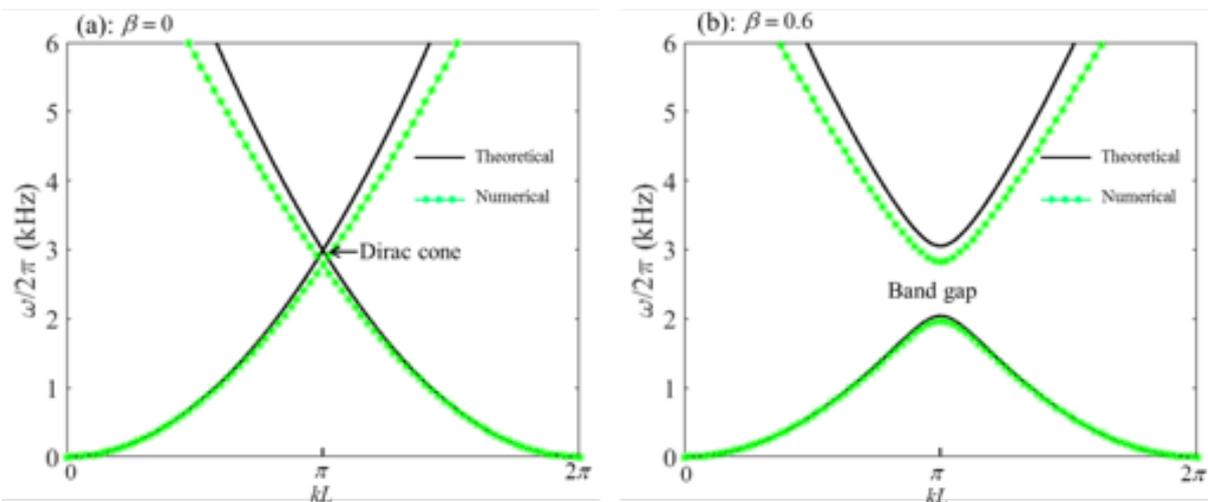

Fig. 4. The band structure of the active sandwiched beam system for $\gamma = 1.19$. (a) $\beta = 0$; (b) $\beta = 0.6$.

The mapping of band gap as a function of $\beta$ for $\gamma = 1.19$ is shown in Fig. 5, with $\beta$ in the range $-0.7 \leq \beta \leq 0.7$. According to the discussion of Fig. 3, we know that $\beta$ in this range does not induce instability. In the figure, the blue (red) dashed (solid) curve denotes the theoretical band gap frequency of the antisymmetric (symmetric) edge mode, while the black solid curves with asterisks are the numerical frequencies of edge modes. The mode shapes of these symmetric and antisymmetric edge modes are presented on the right of the figure. Similar to Fig. 4, the numerical edge frequencies are lower than their theoretical counterparts. From the figure we find that the band gap is closed first for increasing $\beta$ from $\beta = -0.7$ to $\beta = 0$, while for further increases in $\beta$, the band gap reopens. Such a process of closing and reopening of band gap is accompanied by a mode exchange: for $\beta < 0$, the lower edge mode is symmetric and the higher edge mode is antisymmetric; while for $\beta > 0$, these two edge modes are exchanged with each other. According to our knowledge of topological properties [33, 34], the mode exchange may cause topological phase transition, thus making the band gaps on the two sides of the Dirac point (i.e. the point when the band gap in closed, here when $\beta = 0$) to have different topological properties. To check whether the topological property of the band gap is changed during the mode exchange, we need to determine the Zak phase of the neighboring pass bands.

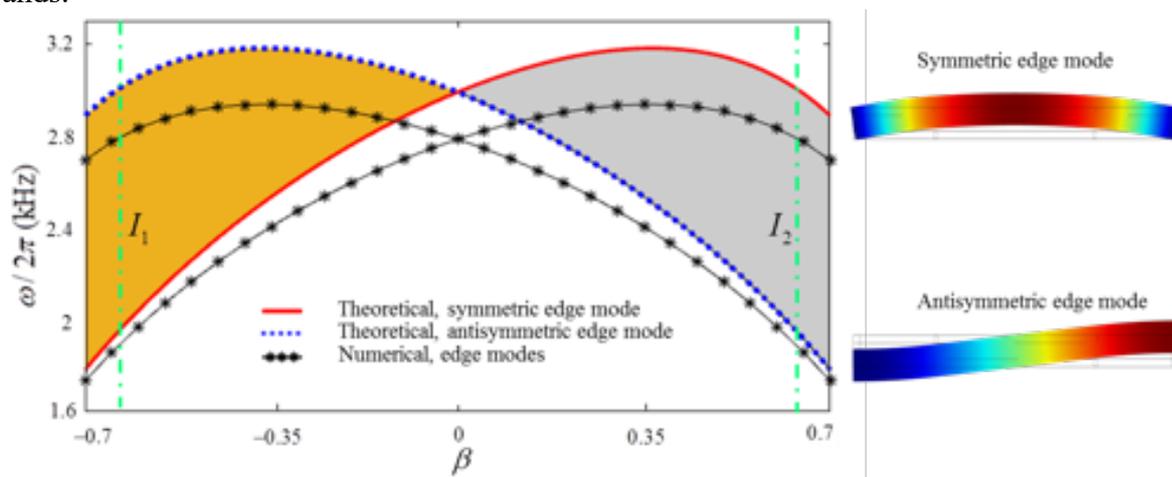

Fig. 5. The mapping of band gap as a function of $\beta$.

The Zak phase is an important topological invariant that determines the topological performance of 1D PCs. The Zak phase is nothing but an integration of the Berry connection over the first Brillouin zone, expressed as

$$\theta_n^{Zak} = i \oint dk \int_\Omega \left( u^* \frac{\partial}{\partial k} u \right) dv \qquad (19)$$

where the subscript $n$ denotes the $n$-th pass band, $u$ is the periodic-in-cell mode function, i.e. $u = e^{-ikL} w$ with $w$ being the normal mode function, the superscript * denotes conjugation, and $\Omega$ denotes the unit cell volume. Generally, there are two ways, by qualitative and quantitative methods, to determine the Zak phase. The quantitative method refers to a numerical analysis based on the definition of the Zak phase $\theta_n^{Zak}$ given in Eq. (19). However, the direct use of definition to determine Zak phase is rather difficult, so its approximated discrete form is commonly preferred [33]. The qualitative method is based on analyzing the mode shapes at $kL = 0$ and at $kL = \pi$ [52, 53]. For the $n$-th pass band, when both the modes at $kL = 0$ and $kL = \pi$ points are symmetric (or both antisymmetric), we can predict a zero Zak phase $\theta_n^{Zak}$; otherwise $\theta_n^{Zak} = \pi$ [33]. Here the qualitative method to predict the Zak phase is adopted. For $\beta < 0$ (take state $I_1$ as an example, $\beta = -0.6$), both the modes of the first band at $kL = 0$ and $kL = \pi$ are symmetric, and both the modes of the second band at $kL = 0$ and $kL = \pi$ are antisymmetric. Therefore, for $\beta < 0$, it is predicted that $\theta_1^{Zak} = 0$ and $\theta_2^{Zak} = 0$. However, for $\beta > 0$ (take state $I_2$ as an example, $\beta = 0.6$), the modes of the first band and the second band are interchanged at $kL = \pi$, while they remain unchanged at $kL = 0$. Thus, it is determined that $\theta_1^{Zak} = \pi$ and $\theta_2^{Zak} = \pi$. In conclusion, the Zak phases of the first two bands for $\beta < 0$ are different with respect to that for $\beta > 0$. This phenomenon proves that the topological property of the band gap for $\beta < 0$ (the yellow domain in the figure) is different from that of the band gap for $\beta > 0$ (the gray domain in the figure).

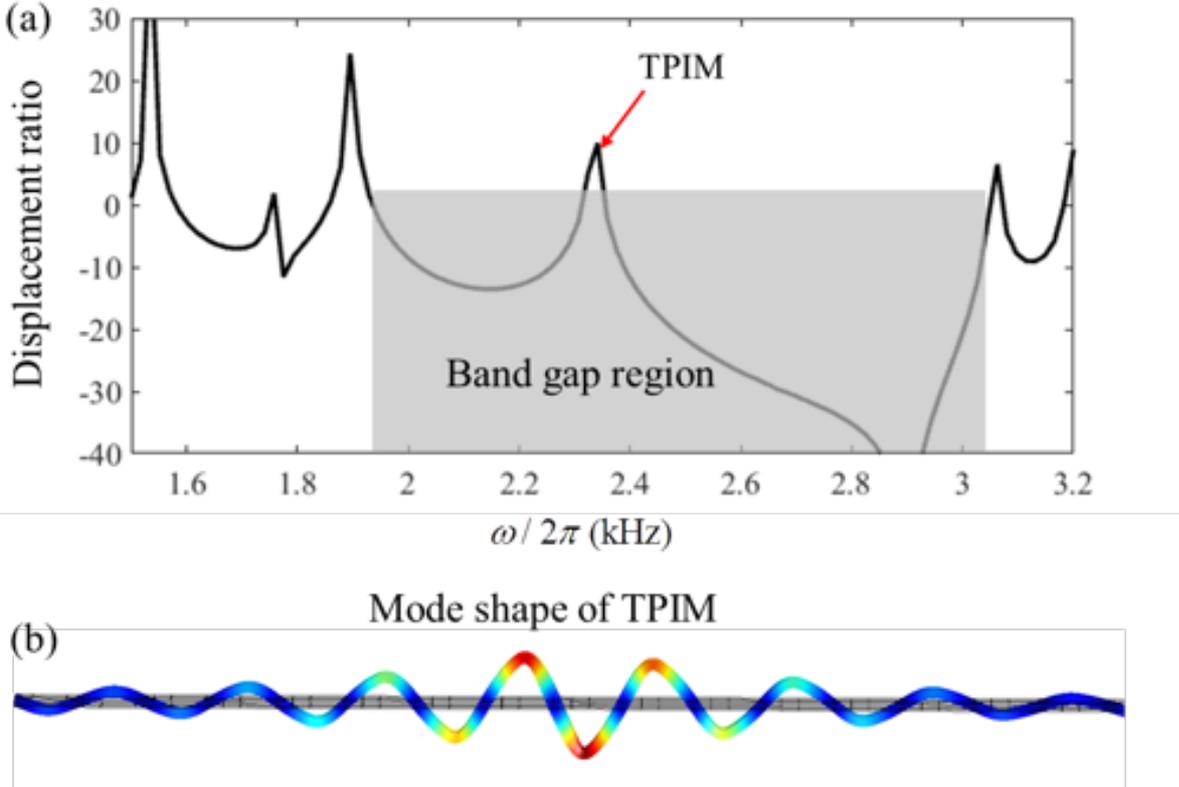

Fig. 6. Transmission of the flexural wave in the composite beam system constructed by cells of state $I_1$ and $I_2$. A hallmark in this composite system is the existence of TPIM.

It is well understood that there may be a TPIM located at an interface between two PC systems with band gaps (sharing an overlap region) of different topological properties. Thus, we design a composite beam system that is constructed by 8 beam cells of state $I_1$ and 8 beam cells of state $I_2$.

The right end of the composite beam system is fixed and on the left end there is a harmonic excitation. The system output is received at the interface. We define the displacement ratio as

$$\alpha = 20\log_{10}\left(\frac{|u_{output}|}{|u_{input}|}\right) \quad (20)$$

The relation of displacement ratio and excitation frequency is shown in Fig. 6(a) with the gray domain showing the band gaps of states $I_1$ or $I_2$. We observe that within the band gap region, there is a transmission peak that stands for the TPIM. The mode shape of TPIM is shown in Fig. 6(b). We can see that the vibration energy is concentrated around the interface. We define a quality factor $\eta$ to describe energy concentration around the interface as

$$\eta = \frac{\int_{\Omega_I} |\mathbf{u}|^2 \, dv}{\int_{\Omega_T} |\mathbf{u}|^2 \, dv} \quad (21)$$

where $\mathbf{u}$ is displacement field, $\Omega_I$ stands for the two unit cells neighboring to the interface while $\Omega_T$ denotes the whole structure. For the structure and physical parameters considered here, the quality factor is $\eta = 0.393$, which means that 39.3% of energy concentrates around the interface.

As shown in Fig. 6, TPIM exists at a single frequency point. In practical applications, such TPIM with very narrow working frequency domain is hard to use because the variation of environmental factors, ie. environmental temperature and water pressure, may change the structural stiffness that in turn will shift the TPIM location, making the designed devices based on TPIM fail. Thus, it is important to broaden the working range of TPIM. Here we show a method to broaden the TPIM working range by introducing an active control operation such that the TPIM working frequency can be tuned at a wider range.

We now consider a composite beam system that is constructed by 8 beam cells with $\beta = \beta_I$ and another 8 cells with $\beta = -\beta_I$. The referential negative capacitance $C_{N0}$ is set as a constant ($\gamma = 1.19$). The TPIM frequency and quality factor in the composite system as functions of parameter $\beta_I$ are displayed in Figs. 7 (a) and (b), respectively. For decreasing $\beta_I$ from $\beta_I = 0.7$ to $\beta_I = 0.2$, the TPIM frequency can be changed widely, from 2.12 kHz to 2.79 kHz, at the cost of a decrease in the quality factor, from $\eta = 0.485$ to $\eta = 0.206$. For reference, the mode shapes of TPIMs are shown in Fig. 7(c) for $\beta_I = 0.3$, 0.5 or 0.7. It is clear that for $\beta_I = 0.5$ or $\beta_I = 0.7$, the TPIM quality is relatively good because of very weak vibration when the component is far away from the interface. However, the TPIM quality for $\beta_I = 0.3$ is poor because the vibration of components far from the interface is relatively strong, and consequently energy concentration is not good. Further, by selecting TPIMs with quality factor $\eta$ larger than $\eta = 0.35$, from Fig. 7(b), we observe that $\beta_I$ should be in the range $\beta_I > 0.46$. In turn, from Fig. 7(a), we find that the TPIM frequency with quality factor larger than 0.35 can be varied from 2.13 kHz to 2.55 kHz.

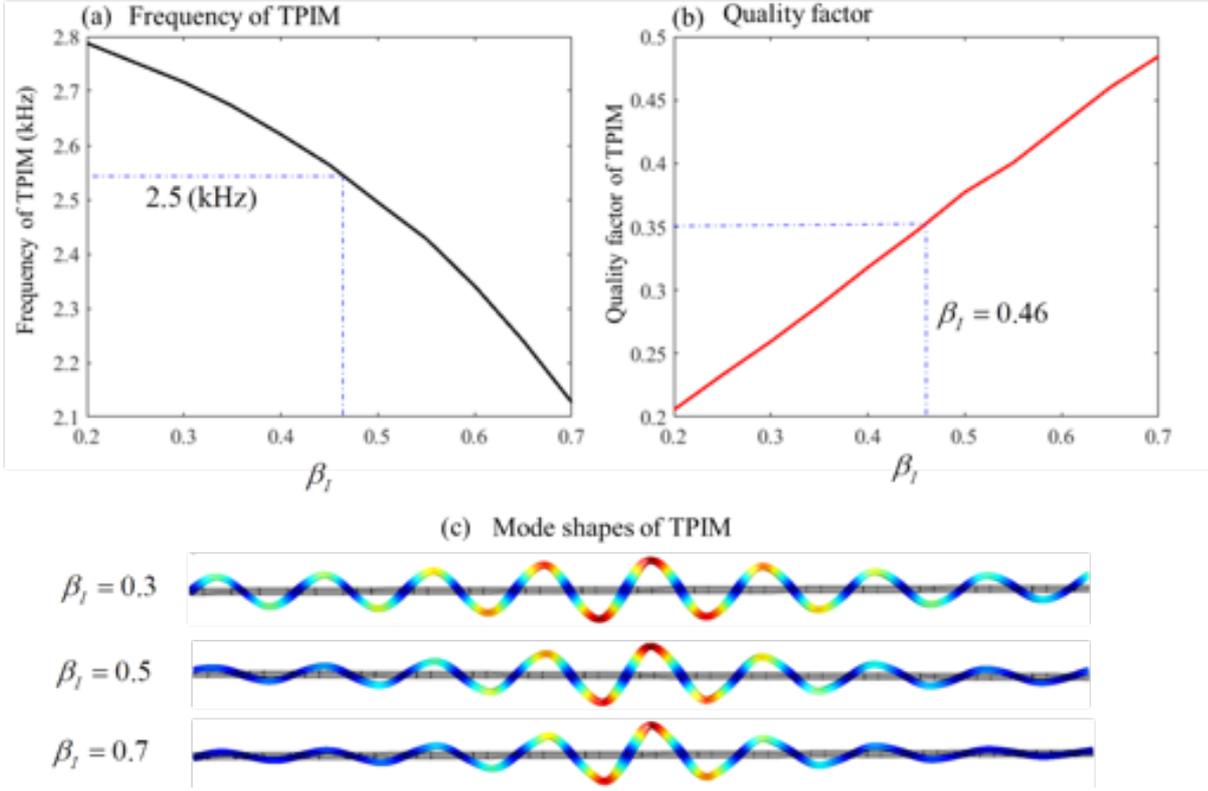

Fig. 7. Active control of TPIM frequency and quality by varying $\beta_I$.

It is shown in Fig. 3(b) that the flexural rigidity of the sandwiched beam relies sensitively on the referential capacitance constant $\gamma$. It is expected that varying $\gamma$ can also significantly change the band gap location in the PC beam system. In Fig. 8(a), the band gap mapping of the pure sandwiched PC beam for $\beta=0.3$, $\beta=0.5$ and $\beta=0.7$ as functions of the parameter $\gamma$ are presented. For $\beta_I$ as a constant, we can obtain the allowed range of $\gamma$ from the analysis in the prior section and Fig. 3(b) as $\gamma_{\min} \leq \gamma \leq \gamma_{\max}$, where

$$\gamma_{\min} = \frac{2.04abI_P d_{31}^2 C_{P2}^{\sigma} + 2abd_{31}^2 I_P \left(C_{P2}^S - 1.02 C_{P2}^{\sigma}\right)\beta_I + h_P s_{11}^E \left(2I_P + E_C I_C s_{11}^E\right)\left(C_{P2}^S - 1.02 C_{P2}^{\sigma}\right)\beta_I C_{P2}^S}{h_P s_{11}^E \left(2I_P + E_C I_C s_{11}^E\right)\left(C_{P2}^S - 1.02 C_{P2}^{\sigma}\right)\beta_I C_{P2}^{\sigma} + 2abI_P d_{31}^2 C_{P2}^{\sigma}}.$$
(22)
$$\gamma_{\max} = \frac{2ab(1+\beta_I)d_{31}^2 I_P}{\beta_I h_P s_{11}^E \left(2I_P + E_C I_C s_{11}^E\right) C_{P2}^{\sigma}} + \frac{C_{P2}^S}{C_{P2}^{\sigma}}.$$

For $\beta=0.3$ and from Eq. (22), we have $\gamma_{\min} = 1.0312$ and $\gamma_{\max} = 1.4976$. For $\gamma$ increasing from $\gamma_{\min}$ to $\gamma_{\max}$, we can see a very wide variation of band gap from the initial range $[1.458(\text{kHz}), 1.792(\text{kHz})]$ to the range $[3.061(\text{kHz}), 3.715(\text{kHz})]$. Similar phenomena can be observed for $\beta=0.5$ and $\beta=0.7$, except that the band gap tunable ranges are narrower compared to $\beta=0.3$. A comparison between these three situations leads to the conclusion that a larger $\beta$ results in a narrower band gap controllable range. The dependence of band gap on $\gamma$ proves that it is also efficient to control the TPIM behavior via the variation of $\gamma$.

Now consider a composite beam system composed of 8 beam cells with $\beta = \beta_I$ and another 8 beam cells with $\beta = -\beta_I$. A constant $\beta_I$ is set and the dependence of the TPIM location and quality on $\gamma$ is investigated. In Fig. 8(b), we observe that $\gamma$ affects sensitively the TPIM frequency. For $\beta_I = 0.3$, its frequency can be tuned from 1.6 (kHz) to 3.11 (kHz), which is a super wide control range. Besides, for $\beta_I = 0.5$ the TPIM control ranges from 1.74 (kHz) to 2.7 (kHz) and for $\beta_I = 0.7$, the control ranges from 1.85 (kHz) to 2.17 (kHz). Obviously, from the viewpoint of control range, $\beta_I = 0.3$ is the best among these cases. However, from Fig. 8(c) we find that the TPIM quality for $\beta_I = 0.3$ is poor. We also observe that the TPIM quality factor depends mainly on $\beta_I$. For a constant $\beta_I$, varying $\gamma$ has little influence on the quality behavior. Therefore, we are able to widely tune the TPIM frequency and at the same time to keep its concentration behavior. In addition, a conflict between control range and quality factor is observed in Figs. 8(b) and 8(c). A wide TPIM control range is at the cost of a bad concentration behavior, while a higher quality factor induces a narrower control range. Consequently, we conclude that the TPIM active control method proposed here is unable to achieve high quality and wide control range simultaneously.

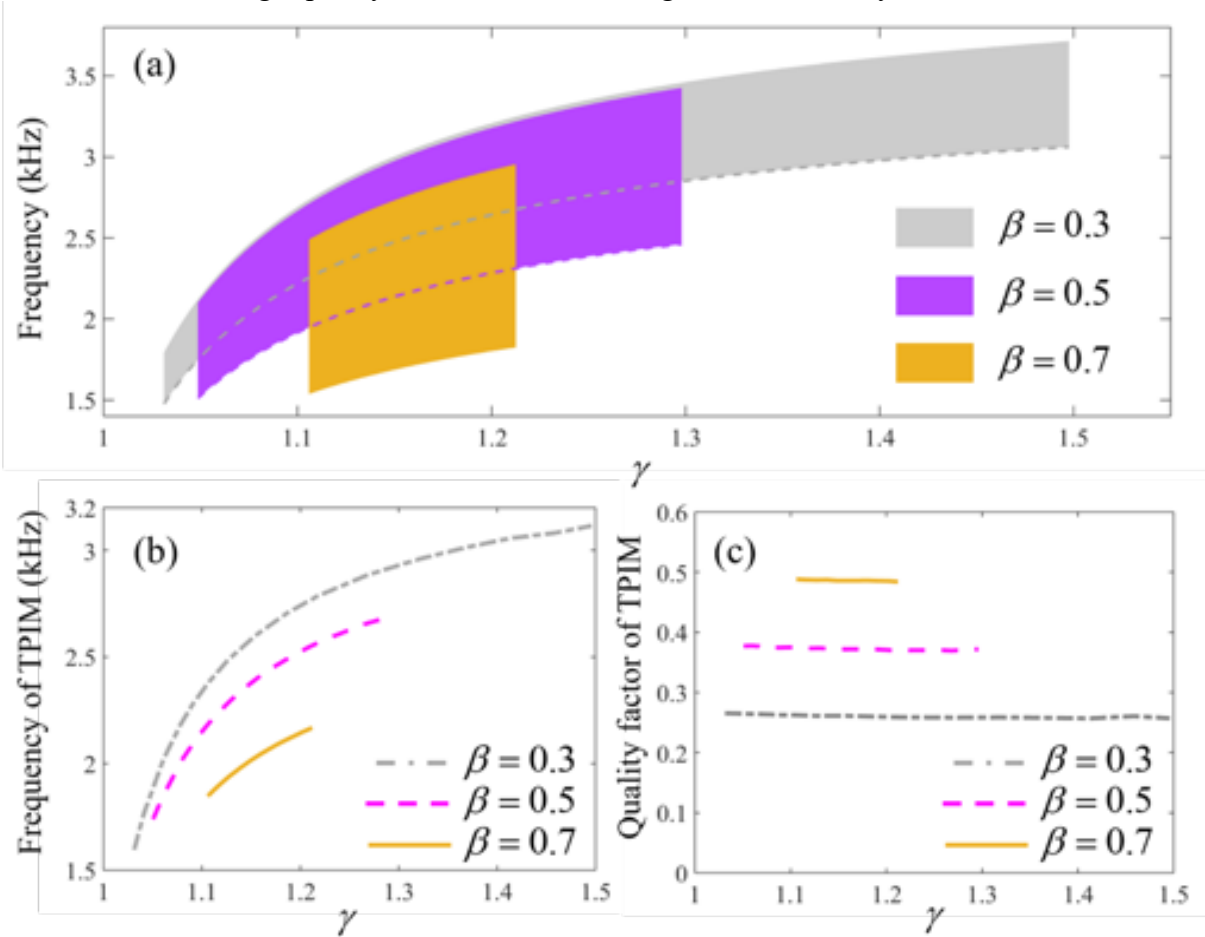

Fig. 8. The active control of the TPIM via variation of parameter $\gamma$.

Although an experimental study of tunable TPIM via negative capacitors is not within the scope of this paper, this aspect of research is certainly important and it will be conducted and reported in the future. A proper guidance for designing and setting up experiments can be concluded from some previous experimental works [47, 54]. However, in the experiments for tunable TPIM, some potential challenges still exist. These include the presence of damping, nonlinear response, and particularly the precise control of negative capacitors because it is sensitive to temperature variation when its absolute value is close to the capacitor of the piezoelectric beam. These factors will be comprehensively studied in the future to realize the application of tunable topological insulators.

## 4. Conclusion

An innovative active composite beam made of epoxy for the central beam and sandwiched by piezoelectric beams is proposed in this paper. The piezoelectric beam surfaces are connected to periodic NCs. We observe a topological phase transition in this system when the values of these NCs are changed in a certain way.

Although it is known that TPIM exists in an interface between two PC systems of different topological properties, usually these TPIMs only locate at a single frequency point. Due to environmental factors including temperature, mechanical loading, etc., that affects the device properties, practical applications of the devices are extremely difficult. To solve this problem, we propose an innovative method to actively and smoothly tune the TPIM frequency such that the resulting TPIM devices are intelligent and robust against the environmental perturbations. An active composite beam system, that is formed by combining two active PC beams with opposite $\beta$, is designed and TPIM can be observed at the interface. Numerical simulations show that for a constant $\gamma$, the variation of $\beta$ can widely change the TPIM location, and further influences its quality factor. On the other hand, for a fixed $\beta$, varying $\gamma$ can also widely change the TPIM frequency while keeping its quality factor almost unchanged.

It should be pointed out that the parameters $\beta$ and $\gamma$ are not arbitrary. They are constrained by the electrical stability and the mechanical stability of the system. Based on the stability analyses of the NC, we obtain the theoretical expressions of the allowed ranges of the parameters $\beta$ and $\gamma$. The NC values considered in this paper are all in the stable region.

## Acknowledgments

The work described in this paper was supported by PROCORE-France/Hong Kong Joint Research Scheme (Project No. FCityU108/17) and National Natural Science Foundation of China (Nos. 11532001 and 11621062).